\documentclass{iau}
\usepackage{graphicx}

\title[A Brief History of NYMGs] %% give here short title %%
{A Brief History of the Study of Nearby Young Moving Groups and Their Members}

\author[J. H. Kastner]   %% give here short author list %%
{Joel H. Kastner}

\affiliation{Laboratory for Multiwavelength Astrophysics, Center
  for Imaging Science, and School of Physics and Astronomy, Rochester Institute of Technology; {\tt jhk@cis.rit.edu}
}

\pubyear{2015}
\volume{314}  %% insert here IAU Symposium No.
\pagerange{119--126}
\jname{Young Stars \& Planets Near the Sun}
\editors{J. H. Kastner, B. Stelzer, \& S. A. Metchev, eds.}
\begin{document}

\maketitle

\begin{abstract}
  Beginning with the enigmatic (and now emblematic) TW Hya, 
  the scutiny of individual stars and star-disk systems has both
  motivated and benefitted from the identification of nearby young
  moving groups (NYMGs). I briefly outline the emergence of this
  relatively new subfield of astronomy over the past two decades, and
  offer a few examples illustrating how the study of NYMGs and their members
  enables unique
  investigations of pre-main sequence stellar evolution, evolved
  protoplanetary disks, and young exoplanets. 
\end{abstract}

\firstsection % if your document starts with a section,

\section{TW Hya: a Classical T  Tauri Star without a Birthplace}

The identification and investigation of {\it nearby young moving groups} (NYMGs) ---
i.e., loose associations of stars that lie within $\sim$100 pc of Earth
and have ages up to $\sim$100 Myr --- arguably began with the identification of the late-type, emission-line field star TW Hydrae
as a ``T Tauri star far from any dark cloud'' \cite[(Rucinski \&
Krautter 1983)]{1983A&A...121..217R}. This comprehensive study by Rucinski \& Krautter establishing the classical T Tauri nature of TW Hya was motivated by the work of \cite[Henize (1976)]{H76} and \cite[Herbig (1978)]{H78}, and led directly to the identification of additional examples of ``isolated T Tauri stars'' \cite[(de la Reza et al.\ 1989; Gregorio-Hetem et al.\ 1992)]{dlR89,GH92}. 
These studies yielded the identification of a literal handful of late-type field stars with T Tauri-like characteristics --- in particular, deep Li absorption and apparent mid-infrared excesses\footnote{It is fascinating to note, in retrospect, that two of the five original members of the TW Hya Association were only observed in H$\alpha$ and Li, and hence identified as young, by \cite[de la Reza et al.\ (1989)]{dlR89} and \cite[Gregorio-Hetem et al.\ (1992)]{GH92} thanks to chance alignments with IR-bright background galaxies.} --- in the general vicinity of TW Hya. 

In their landmark paper, \cite[de la Reza et al.\ (1989)]{dlR89} speculated on the potential nature of these lonely T Tauri stars, suggesting they were ``kinematically related'' and likely formed {\it in situ}. Furthermore, on the basis of these stars' high galactic latitudes, de la Reza et al.\ suggested (without assigning a specific distance range) that TW Hya and its apparent cohort may be relatively close to Earth. The last assertion seemed particularly intriguing, given the potential for ``close-up'' studies of planetary systems in formation that might be afforded by these and other ``isolated'' T Tauri stars. Following up on this hunch, \cite[Zuckerman et al.\ (1995)]{Z95} established that the thermal infrared excess from dust around TW Hya (and a few other ``isolated,'' young, IR-excess stars) was accompanied by the presence of residual orbiting molecular (CO) gas, suggestive of the recent cessation of an epoch of planet formation. But the distance and age of TW Hya remained very poorly constrained. 

With the release of the ROSAT (X-ray) All-Sky Survey (RASS), we could address these unknowns. The RASS X-ray data firmly established the youth of TW Hya and its (four) nearby siblings, yet also suggested the five were older than typical of T Tauri stars in dark clouds; we guesstimated (sic) the ages of the five stars as $\sim$20 Myr \cite[(Kastner et al.\ 1997)]{K97}. With this age estimate, we could then place firmer constraints on the stars' distances, and determined that they were a mere $\sim$40--60 pc from Earth. This range was promptly confirmed by the newly released Hipparcos parallax distances to TW Hya ($D=54$ pc) and one of its four siblings, HD 98800 ($D=45$ pc). In light of their similar ages, distances, and kinematics, we took the (perhaps unwise) leap of christening this little ragtag group of young stars the TW Hya Association (TWA), and declared the TWA ``the nearest region of recent star formation'' \cite[(Kastner et al.\ 1997)]{K97}. 

Fortunately for us, many subsequent studies --- beginning with \cite[Webb et  al.\ (1999)]{Webb99}, and extending to the work of Murphy et al.\ presented at this meeting --- have dramatically expanded the TWA's known membership, have confirmed its mean distance as $\sim$50 pc, and have honed estimates of its age, which is now generally accepted as $\sim$8 Myr\footnote{This (traceback-based) age estimate seems relatively robust (\cite[Ducourant et al.\ 2014]{Duc14}). However, given the many important caveats regarding ages of NYMGs that have been raised at this meeting (see, e.g., reviews by Mamajek and Bell), the age of the TWA certainly bears revisiting via other methods.}. Meanwhile, TW Hya itself has gone on to become the Crab Nebula\footnote{With apologies, and thanks, to David Wilner.\hspace{2.5in} } of late-stage pre-main sequence accretion and protoplanetary disk studies, with nearly 1000 \verb+simbad+ references (95\% of them since 1997) as of the writing of this review.

\section{The (young association) link between the TWA and $\beta$ Pic}

Not long after the naming of the TWA, \cite[Jura et al.\ (1998)]{J98} noted that the A stars HR 4796A, $\beta$ Pic, and 49 Cet --- all massive debris disk hosts --- were underluminous for their colors relative to the field A star population, suggesting all three are young. The youth of the HR 4796AB system had already been established by \cite[Stauffer et al.\ (1995)]{Stauff95} on the basis of the position of the M-type star HR 4796B {\it above} the main sequence. Based on the (relative) sky proximity of HR 4796AB to the original five stars of the TWA, \cite[Jura et al.\ (1998)]{J98} further speculated that HR 4796A may also be a TWA member (an assertion subsequently confirmed by  \cite[Webb et  al.\ 1999]{Webb99}). 

A strong implication of the \cite[Jura et al.\ (1998)]{J98} analysis was that $\beta$ Pic and the stars of the TWA, though not kinematically or spatially associated, were similarly young. At  the time, this was somewhat of a revelation, given that some previous estimates had put the age of $\beta$ Pic at a few 100 Myr (\cite[Barrado y Navascu{\'e}s et al.\ 1999]{1999ApJ...520L.123B} and refs.\ therein). The youth of $\beta$ Pic was soon firmly established by \cite[Barrado y Navascu{\'e}s et al.\ (1999)]{1999ApJ...520L.123B}, who honed lists of stars potentially comoving with $\beta$ Pic and identified just three M star ``survivors,'' all of age $\sim$20 Myr. This work constituted the first identification of the $\beta$ Pic Moving Group ($\beta$PMG) --- its original membership, like that of the original TWA, could be counted on one hand. Not long thereafter, the wider availability and improved application of space velocity data would begin to dramatically expand the known membership of the $\beta$PMG \cite[(Zuckerman et al.\ 2001)]{2001ApJ...562L..87Z}. 

At  the risk of stating the obvious, the general approach pioneered by Zuckerman (e.g., \cite[Zuckerman \& Webb 2000]{2000ApJ...535..959Z}; \cite[Zuckerman et al.\ 2001]{2001ApJ...562L..87Z}), Song (e.g., \cite[Song et al.\ 2002]{2002A&A...385..862S}), Torres \cite[(Torres et al.\ 2006)]{2006A&A...460..695T}, and a few other investigators --- i.e., selecting candidate young stars on the basis of proper motions and/or X-ray fluxes, and following up spectroscopically to confirm signatures of youth and obtain $UVW$ velocities --- has been refined and improved tremendously over the past $\sim$15 years. In particular, statistically rigorous treatments of stellar kinematics (e.g., \cite[Malo et al. 2013]{2013ApJ...762...88M}) have replaced the original, painstaking, ``by-hand'' searches of the Hipparcos and Tycho catalogs, and the Galex UV sky survey data have replaced the RASS as a superior means to select active (hence, possibly young) late-type stars in the field (\cite[Shkolnik et al.\ 2011]{2011ApJ...727....6S}; \cite[Rodriguez et al.\ 2013]{2013ApJ...774..101R}). Thus, the techniques first brought to bear to identify and confirm members of the TWA and $\beta$PMG have now been successfully applied to identify (and/or expand the memberships of) another dozen or so established and candidate NYMGs (see, e.g., \cite[Zuckerman et al.\ 2011]{2011ApJ...732...61Z} and review by Mamajek in these Proceedings).

\section{The impact of nearby young moving group studies}

I summarize here three recent examples of studies involving $\beta$PMG members that highlight the utility of studies of NYMGs and their members for purposes of improving our understanding of pre-main sequence (pre-MS) stellar evolution, evolved protoplanetary disks, and young exoplanets.

\subsection{Understanding early stellar evolution}

The comprehensive study of the $\beta$PMG by \cite[Mamajek \& Bell (2014)]{2014MNRAS.445.2169M} well illustrates how a NYMG can serve as a testbed both for the astrophysics of pre-main sequence stars and techniques for age-dating stars and stellar associations. Mamajek \& Bell compiled and critically compared a dozen $\beta$PMG age determinations involving roughly a half-dozen different techniques, including placement on pre-MS star isochrones, kinematic traceback, Li depletion, and variants thereof. They conclude that the age of the $\beta$PMG should be revised upwards, from the widely quoted $\sim$12 Myr (\cite[Zuckerman et al.\ 2001]{2001ApJ...562L..87Z}; \cite[Torres et  al.\ 2006]{2006A&A...460..695T}) to $\sim$23 Myr (which is, ironically, closer to the original estimate by \cite[Barrado y Navascu{\'e}s et al.\ 1999]{1999ApJ...520L.123B}). However, this refinement in the age of the $\beta$PMG is perhaps less interesting than the conclusion by \cite[Mamajek \& Bell (2014)]{2014MNRAS.445.2169M} that, at  least in application to the $\beta$PMG stars, the Li depletion boundary  and isochronal age estimation techniques are superior to kinematic methods of age determination. Given the interesting, fundamental problems in stellar astrophysics that are inherent in the former two methods (as highlighted in talks by Bell, Song, Somers and others at this meeting), it is clear that the tests of Li depletion boundary  and isochronal age determinations offered by studies of NYMGs will become increasingly important, moving forward.

\subsection{Investigating (evolved) protoplanetary disks}

\begin{figure}[h!]
\begin{center}
\includegraphics[width=5.5in]{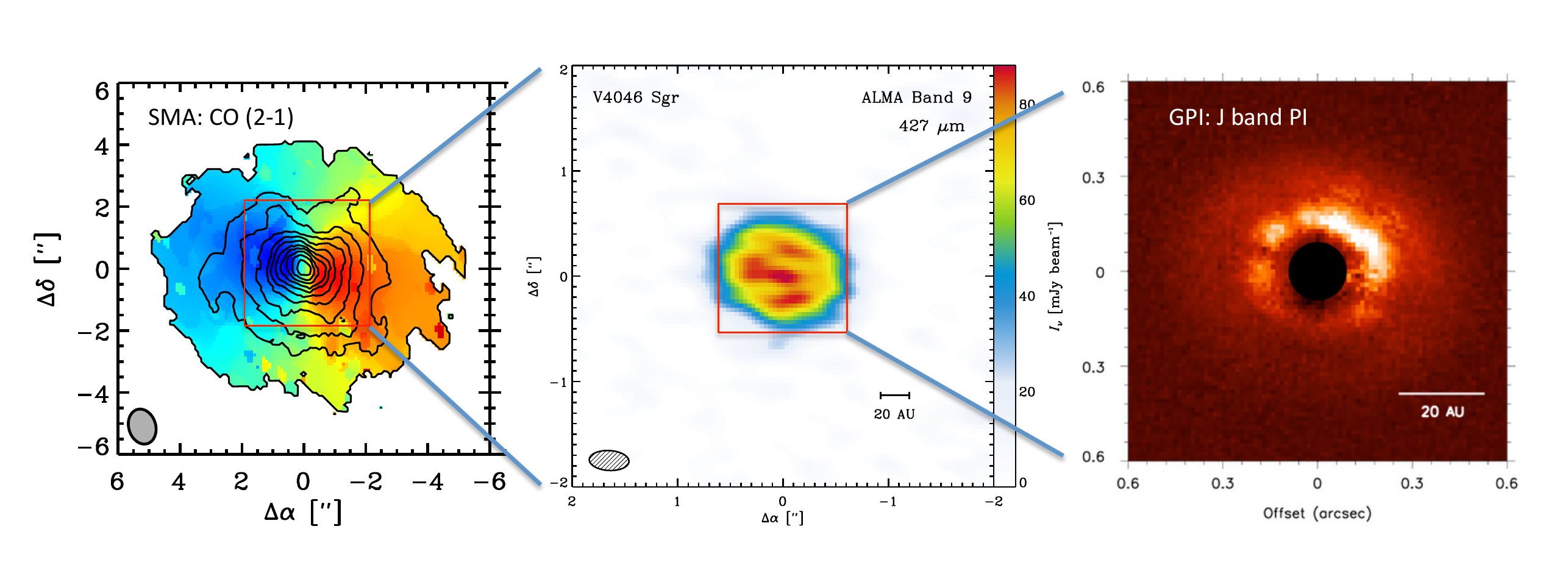}
\caption{ The V4046 Sgr disk as imaged in CO line emission by the Submillimeter Array (left; from \cite[Rosenfeld et al.\ 2013]{2013ApJ...775..136R}), in 427 $\mu$m continuum emission by ALMA during Early Science (Cycle 0) operations (center; from Andrews et al., in prep.), and in polarized intensity at 1.25 $\mu$m by the Gemini Planet Imager in its coronagraphic/polarimetric mode (right;  from \cite[Rapson et al.\ 2015]{2015ApJ...803L..10R}).  
%Bottom row: simulation of a protoplanetary disk viewed at i =45? with a gap cleared by a single Jupiter-mass planet (located at the position of the X), as seen in synthetic submm (left) and near-IR (right) images, from Dong et al (2015).
}
\label{fig:V4046SgrModels}
\end{center}
\end{figure}

Though it remains understudied, especially relative to TW Hya, the $\beta$PMG member system V4046 Sgr is nearly as close ($D \sim 73$ pc) and, arguably, at least as interesting\footnote{For a brief overview of V4046 Sgr, see the May 2013 Star Formation Newsletter:
\\ http://www.ifa.hawaii.edu/$\sim$reipurth/newsletter/newsletter245.pdf.}. This hierarchical multiple system features a close, near-equal-components, solar-mass binary that is orbited by, and actively accreting from, a molecule-rich circumbinary disk. The disk is surprisingly extensive and massive (outer CO radius $\sim$350 AU and estimated H$_2$ mass $\sim$0.09 $M_\odot$, \cite[Rosenfeld et al.\ 2013]{2013ApJ...775..136R}; Fig.~\ref{fig:V4046SgrModels}, left panel) given the advanced age of V4046 Sgr and, hence, potentially affords the opportunity to study late stages in the process of planet formation around a close binary. Recently, \cite[Rapson et al.\ (2015)]{2015ApJ...803L..10R} used the new Gemini Planet Imager (GPI) to obtain near-infared coronagraphic/polarimetric (scattered-starlight) images of the innermost disk regions (Fig.~\ref{fig:V4046SgrModels}, right panel). Thanks to the combination of GPI's exquisite performance and the proximity of V4046 Sgr, these images probe the structure of the dust disk to within $\sim$7 AU of the central binary, at a jaw-dropping $\sim$3 AU resolution. The GPI imaging reveals concentric rings of submicron-sized dust grains that are contained within the ring of submm-sized grains responsible for thermal emission in submm continuum imaging. The structures seen in these (GPI and ALMA) images of V4046 Sgr (Fig.~\ref{fig:V4046SgrModels}) are strikingly similar to those seen in synthetic images derived from simulations describing newborn planets clearing gaps within a protoplanetary disk (see, e.g., Fig.~7 in \cite[Dong et al.\ 2015]{2015ApJ...809...93D}).

\subsection{Discovering and characterizing young exoplanets}

Because young gas giant planets are self-luminous in the
infrared, NYMG members offer the best targets for 
extreme adaptive optics imaging of giant exoplanets. Indeed, direct imaging of nearby,
young stars represents the {\it only} means that will be available
in the near future to detect exoplanets in wide
($>$5 AU) orbits around their host stars (see, e.g., review by Chauvin in these Proceedings). 
This potential for NYMG members --- especially those hosting dusty debris disks --- to yield direct-imaging detections of massive exoplanets is, of course, amply demonstrated by the detection and subsequent intensive study of $\beta$ Pic b (e.g., \cite[Bonnefoy et 
al.\ 2014]{2014A&A...567L...9B}). With GPI and SPHERE now having demonstrated their capabilities, we appear to be on the verge of a trove of such direct-imaging discoveries of planets orbiting NYMG member stars.

\acknowledgments{JHK's research on young stars near Earth is supported by NSF grant AST 1108950 and NASA Astrophysics Data Analysis Program grant NNX12AH37G to RIT.}

\end{document}